\documentclass[aps,pra,twocolumn,showpacs,amsmath,amssymb,superscriptaddress]{revtex4}
\usepackage{amsfonts}
\usepackage[dvips]{graphics,color}

\newcommand{\msc}[1]{{\scriptstyle #1}}
\newcommand{\mscs}[1]{{\scriptscriptstyle #1}}
\newtheorem{theorem}{Theorem}
\newtheorem{lemma}{Lemma}

\begin{document}

\title{Non-local correlations as an information theoretic resource}

\author{Jonathan Barrett}
\email{jbarrett@ulb.ac.be}
\affiliation{Physique Th\'{e}orique, {C.P.} 225, Universit\'{e} Libre de Bruxelles, Boulevard du Triomphe, 1050 Bruxelles, Belgium}
\affiliation{Centre for Quantum Information and Communication, {C.P.} 165/59, Universit\'{e} Libre de Bruxelles, Avenue F. D. Roosevelt 50, 1050 Bruxelles, Belgium} 
\author{Noah Linden}
\email{n.linden@bristol.ac.uk}
\affiliation{Department of Mathematics, University of Bristol, University Walk, Bristol BS8 1TW, United Kingdom}
\author{Serge Massar}
\email{smassar@ulb.ac.be}
\affiliation{Physique Th\'{e}orique, {C.P.} 225,
Universit\'{e} Libre de Bruxelles, Boulevard du Triomphe, 1050 Bruxelles, Belgium}
\affiliation{Centre for Quantum Information and Communication, {C.P.} 165/59, Universit\'{e} Libre de Bruxelles, Avenue F. D. Roosevelt 50, 1050 Bruxelles, Belgium} 
\author{Stefano Pironio}
\email{spironio@ulb.ac.be}
\affiliation{Physique Th\'{e}orique, {C.P.} 225,
Universit\'{e} Libre de Bruxelles, Boulevard du Triomphe, 1050 Bruxelles, Belgium}
\affiliation{Centre for Quantum Information and Communication, {C.P.} 165/59, Universit\'{e} Libre de Bruxelles, Avenue F. D. Roosevelt 50, 1050 Bruxelles, Belgium} 
\author{Sandu Popescu}
\email{s.popescu@bristol.ac.uk}
\affiliation{{H.H.} Wills Physics Laboratory, Tyndall Avenue, Bristol BS8 1TL, United Kingdom}
\affiliation{Hewlett-Packard Laboratories, Stoke Gifford, Bristol BS12 6QZ, United Kingdom}
\author{David Roberts}
\email{david.roberts@bris.ac.uk}
\affiliation{{H.H.} Wills Physics Laboratory, Tyndall Avenue, Bristol BS8 1TL, United Kingdom}

\begin{abstract}
It is well known that measurements performed on spatially separated 
entangled quantum systems can give rise to correlations that are non-local, 
in the sense that a Bell inequality is violated. They cannot, however, be 
used for super-luminal signalling. It is also known that it is possible to 
write down sets of ``super-quantum'' correlations that are more non-local 
than is allowed by quantum mechanics, yet are still non-signalling. Viewed 
as an information theoretic resource, super-quantum correlations are very 
powerful at reducing the amount of communication needed for distributed 
computational tasks.  An intriguing question is why quantum mechanics does 
not allow these more powerful correlations.
We aim to shed light on the range of quantum possibilities by placing them 
within a wider context. With this in mind, we investigate the set of 
correlations that are constrained only by the no-signalling principle. 
These correlations form a polytope, which contains the quantum correlations 
as a (proper) subset. We determine the vertices of the no-signalling 
polytope in the case that two observers each choose from two possible 
measurements with $d$ outcomes. We then consider how interconversions 
between different sorts of correlations may be achieved. Finally, we 
consider some multipartite examples.


\end{abstract}

\maketitle

\section{Introduction}

In a typical Bell-type experiment, two entangled particles are produced at a source and move apart to separated observers. Each observer chooses one from a set of possible measurements and obtains some outcome. The joint outcome probabilities are determined by the measurements and the quantum state. 
One of the more striking features of quantum mechanics is that joint outcome probabilities can violate a Bell-type inequality \cite{Bell}, indicating that quantum mechanics is not, in Bell's terminology, locally causal.
This prediction has been confirmed in numerous laboratory experiments \cite{aspect}. 

Abstractly this scenario may be described by saying that the two observers have access to a black box. Each observer selects an input from a range of possibilities and obtains an output. The box determines a joint probability for each output pair given each input pair. It is clear that a quantum state provides a particular example of such a box, with input corresponding to measurement choice and output to measurement outcome. More generally, boxes can be divided into different types. Some will allow the observers to signal to one another via their choice of input, and correspond to two-way classical channels, as introduced by Shannon \cite{shannon}. Others will not allow signalling - it is well known, for example, that any box corresponding to an entangled quantum state will not. This is necessary for compatibility between quantum mechanics and special relativity. Of the non-signalling boxes, some will violate a Bell-type inequality. The significance of this can be spelt out in information theoretic terms: separated observers without the box, who have access to pre-shared classical random data but no other resources, and in particular who cannot communicate, will not be able to simulate the box. We refer to any such box (and to the corresponding correlations) as non-local. 

In general, these boxes can be viewed as an information theoretic resource. This is obvious in the case of signalling boxes, or classical channels. However, it is also known that non-local correlations arising from an entangled quantum state, even though they cannot be used directly for signalling, can be useful in reducing the amount of signalling that is needed in communication complexity scenarios below what could be achieved with only shared random data \cite{burman}. A local black box is, of course, simply equivalent to some shared random data, which in turn (depending on the precise nature of the problem) may be better than nothing \cite{commcomplex}. 
    
A good question to ask now is, can any set of non-signalling correlations be produced by measurements on some quantum state? The answer, in fact, is no. This was shown by Popescu and Rohrlich \cite{popescu+rohrlich}, who wrote down a set of correlations that return a value of $4$ for the Clauser-Horne-Shimony-Holt (CHSH) expression \cite{chsh}, the maximum value algebraically possible, yet are non-signalling. The maximum quantum value is given by Cirel'son's theorem as $2\sqrt{2}$ \cite{cirel}. These should be compared with the maximum value obtainable by non-communicating classical observers, which is $2$. Popescu and Rohrlich concluded that quantum mechanics is only one of a class of non-local theories consistent with causality. In terms of our boxes, there are some boxes that are non-signalling but are more non-local than is allowed by quantum mechanics. It is interesting to note that from an information theoretic point of view, some of these latter are very powerful. For example, van Dam has shown \cite{vandam} that two observers who have access to a supply of Popescu-Rohrlich-type boxes would be able to solve essentially any two-party communication complexity problem with only a constant number of bits of communication. This should be contrasted with the quantum case, for which it is known that certain communication complexity problems require at least $n$ bits of communication even if unlimited shared entanglement is available \cite{cleve}.

In this work, we investigate the set of non-signalling boxes, considering them as an information theoretic resource. Clearly this set includes those corresponding to measurements on quantum states as a subset. The motivation for studying the wider set is partly that it is interesting for its own sake. This is true even though no correlations other than quantum correlations have so far been observed in Nature. Our findings are preliminary, but it is already clear that the set of non-signalling boxes has interesting structure, and one finds analogies with other information theoretic resources, in particular with the set of entangled quantum states. This work is not, however, purely academic. Another motivation is that a better understanding of the nature of quantum correlations can be gained by placing them in a wider setting. Only in this way, for example, can one hope to answer Popescu and Rohrlich's original question, of why quantum correlations are not more non-local than they are. More generally, a proper understanding of the information theoretic capabilities of quantum mechanics includes an understanding of what cannot be achieved as well as what can. 

This article is organized as follows. In Sec.~\ref{boxes}, we introduce the convex polytope that describes the set of non-signalling correlations. In Sec.~\ref{twosettpoly}, we examine more closely the particular case of correlations involving two possible inputs, obtaining all the vertices of the corresponding polytope. We then consider, in Sec.~\ref{resconvtwosett}, how interconversions between these extreme points may be achieved using local operations. Sec.~\ref{threepartpoly} is devoted to three-party correlations and in Sec.~\ref{env}, we examine how extremal correlations correlate to the environment. We conclude with some open questions in Sec.~\ref{opquest}.

\section{Two party correlations}
\subsection{Definitions}\label{boxes}

\paragraph*{The no-signalling polytope.}
A bipartite correlation box (hereafter, just ``box'') is defined by a set of possible inputs for each of Alice and Bob, a set of possible outputs for each, and a joint probability for each output pair given each input pair. We denote Alice's and Bob's inputs $\msc{X}$ and $\msc{Y}$ respectively, and their outputs $a$ and $b$. The joint probability of getting a pair of outputs given a pair of inputs is $p_{ab|\mscs{XY}}$. Since $p_{ab|\mscs{XY}}$ are probabilities they satisfy positivity,
\begin{equation}\label{pos} 
p_{ab|\mscs{XY}}\geq 0 \qquad \forall\ a,b,\msc{X,Y}
\end{equation}
and normalization,
\begin{equation}\label{norm}
\sum_{a,b} p_{ab|\mscs{XY}}=1 \qquad \forall \ \msc{X,Y}.
\end{equation}
In this work we only consider non-signalling boxes, i.e, we require that Alice
cannot signal to Bob by her choice of $\msc{X}$, and vice versa.
This means that the marginal probabilities $p_{a|\mscs{X}}$ and
$p_{b|\mscs{Y}}$ are independent of $\msc{Y}$ and $\msc{X}$
respectively:
\begin{eqnarray}
\sum_{b}p_{ab|\mscs{X,Y}} &=& \sum_{b}p_{ab|\mscs{X,Y'}} \equiv
p_{a|\mscs{X}}\qquad \forall \ a,\msc{X},\msc{Y},\msc{Y'}\label{nosiga} \\
\sum_{a}p_{ab|\mscs{X,Y}} &=& \sum_{a}p_{ab|\mscs{X',Y}}
\equiv p_{b|\mscs{Y}} \qquad \forall \ b,\msc{Y},\msc{X},\msc{X'}.\label{nosigb}
\end{eqnarray}

A concrete example of a correlation box is an experiment with
two spin-half particles, with the inputs $\msc{X}$ and $\msc{Y}$
labelling Alice's and Bob's analyzer settings and the outputs $a$ and $b$ labelling the experimental outcomes. In a quantum experiment like this one, it is generally the case that the outcome of the measurement is obtained as soon as the measurement is performed. In addition, the entanglement is destroyed after the measurements, so that if the experiment is to be repeated a new entangled state is needed. We define boxes to have the same properties. Alice can select her input at any time and obtains her output immediately, and similarly Bob. There may of course be a time delay between Alice selecting her input and Bob selecting his input, but this makes no difference to the correlations. Further, after a box is used once, it is destroyed and to repeat the experiment a new box is needed.   

We will always consider that the number of possible inputs and outputs is finite. Since the above constraints are all linear, the set of boxes with a given number of inputs and outputs is a polytope, which we denote by $\mathcal{P}$.
It is easy to see that the set is convex - if two boxes each satisfy the constraints, then a probabilistic mixture of them (defined in the obvious manner) will also do so.

\paragraph*{The local polytope.}
In general, the set of non-signalling boxes can be divided into two types, local and non-local. A box is local if and only if it can be simulated by non-communicating observers with only shared randomness as a resource. This means that we can write 
\begin{equation}
p_{ab|\mscs{XY}}=\sum_{\lambda} p_{\lambda}\,p_{a|X}(\lambda)\,p_{b|Y}(\lambda),
\end{equation}
where $\lambda$ is the value of the shared random data and $p_{\lambda}$ is the probability that a particular value of $\lambda$ occurs. We have that $p_{a|X}(\lambda)$ is the probability that Alice outputs $a$ given that the shared random data was $\lambda$ and the input was chosen to be $\msc{X}$, and similarly for $p_{b|Y}(\lambda)$. 

We recall what is known about the set of local boxes (see for instance \cite{Werner, Pitowsky}). This set is itself a convex polytope, with vertices corresponding to local deterministic boxes (all $p_{a|\msc{X}},p_{b|\msc{Y}}$ are 0 or 1). The positivity conditions of Eq.~(\ref{pos}) are trivial facets of this polytope, while non-trivial facets correspond to Bell-type inequalities. Violation of the latter implies that a point lies outside the local polytope, and that the corresponding box is therefore non-local. We denote the local polytope by $\mathcal{L}$.

\begin{figure}
\scalebox{0.67}{\includegraphics{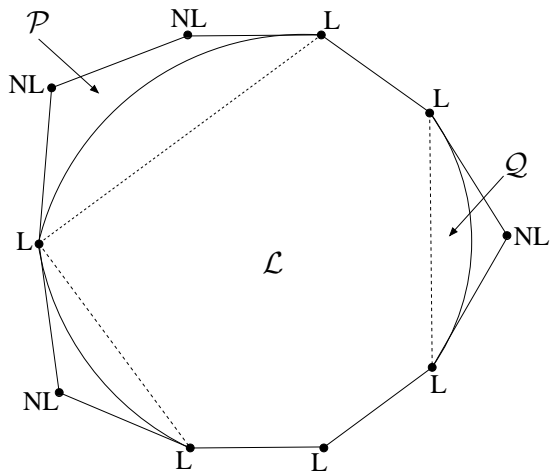}}
\caption[]{A schematic
representation of the space of non-signalling correlation boxes.  The
vertices are labelled L and NL for local and non-local. Bell inequalities are the facets represented in dashed lines. The set bounded by these is $\mathcal{L}$. The
region accessible to quantum mechanics is $\mathcal{Q}$. A general non-signalling
box $\in \mathcal{P}$.\label{nosigpolytopefig}}
\end{figure}

\paragraph*{Quantum mechanical correlations.}
Finally, there is a third set of interest, the correlations obtainable
by measurements on bipartite quantum states. We denote this set $\mathcal{Q}$ (where $\mathcal{Q}$ is defined for a fixed number of measurement settings and outcomes).  The set $\mathcal{Q}$ is investigated in Refs.~\cite{cirel, Cirelson, landau, Pitowsky, masanes}. It is convex but is not a polytope as the number of extremal points is not finite. Since the correlations allowed by quantum mechanics can violate Bell inequalities, $\mathcal{Q}$ is non-local. However, as they violate the CHSH inequality only up to Cirel'son's bound of $2 \sqrt{2}$ \cite{cirel,popescu+rohrlich}, they form a proper subset of the no-signalling polytope. Overall, we have that $\mathcal{L} \subset \mathcal{Q} \subset \mathcal{P}$. This situation is illustrated in Fig.~\ref{nosigpolytopefig}. 

\subsection{The two-inputs no-signalling polytope}\label{twosettpoly}

\subsubsection{Two outputs}\label{twotwotwo}

Having defined the objects that we are interested in, we begin by considering in detail the simple case in which Alice and Bob are each choosing from two inputs, each of which has two possible outputs. We write $\msc{X},\msc{Y},a,b\in\{0,1\}$. The probabilities $p_{ab|\msc{XY}}$ thus form a table with $2^4$ entries, although these are not all independent due to the constraints of Sec.~\ref{boxes}. The dimension of the polytope is found by subtracting the number of independent constraints from $2^4$, and turns out to be 8. To understand the polytope $\mathcal{P}$, we wish to find its
vertices.  These will be boxes that satisfy all of the constraints and saturates a sufficient number of the positivity constraints to be uniquely determined. In the next subsection, we present an argument that allows us to find all the vertices of the two-input $d$-output polytope. Here we simply state the results for the simple two-input two-output case. 

We find that there are 24 vertices, which may be divided
into two classes, those corresponding to local boxes and those corresponding to non-local boxes.
Local vertices are simply the local deterministic boxes, which assign a definite value to each of Alice's and Bob's inputs. There are thus 16 local vertices, which can be expressed as
\begin{equation}
p_{ab|\mscs{XY}} = \left\{ \begin{array}{r@{\quad}c@{\quad}l} 1 &:&
a=\alpha \msc{X}\oplus\beta,\\
&& b=\gamma \msc{Y}\oplus\delta\\
0 &:& \mbox{otherwise,}
\end{array} \right.
\end{equation}
where $\alpha,\, \beta,\,\gamma,\delta \in \{0,1\}$. Here and throughout, $\oplus$ denotes addition modulo 2.

The 8 non-local vertices may be expressed compactly as
\begin{equation}\label{nonlocal}
p_{ab|\mscs{XY}} = \left\{ \begin{array}{r@{\quad: \quad}l} 1/2 &
a\oplus b=\msc{X}.\msc{Y}\oplus\alpha\msc{X}\oplus\beta\msc{Y}\oplus\gamma\\
0 & \mbox{otherwise,}
\end{array} \right.
\end{equation}
where $\alpha,\, \beta,\,\gamma \in \{0,1\}$.  We will refer to these boxes as Popescu-Rohrlich (PR) boxes.

By using reversible local operations Alice and Bob can convert any
vertex in one class into any other vertex within the same class.
There are two types of reversible local operations. Alice may
relabel her inputs, $\msc{X} \rightarrow \msc{X}\oplus 1$, and she may
relabel her outputs (conditionally on the input), $a \rightarrow
a\oplus\alpha\msc{X}\oplus\beta$. Bob can perform similar operations. Thus up to local
reversible transformations, each local vertex is equivalent to
the vertex setting $\alpha=0$, $\beta=0$, $\gamma=0$, $\delta=0$, i.e,
\begin{equation}
p_{ab|\mscs{XY}} = \left\{ \begin{array}{r@{\quad: \quad}l} 1 &
a=0 \,\,\mbox{and} \,\,b=0\\
0 & \mathrm{otherwise.}
\end{array} \right.
\end{equation}
Each non-local vertex is equivalent to
\begin{equation}\label{chshbox}
p_{ab|\mscs{XY}} = \left\{ \begin{array}{r@{\quad: \quad}l} 1/2 &
a\oplus b=\msc{X.Y}\\
0 & \mbox{otherwise.}
\end{array} \right.
\end{equation}
We note that if we allow irreversible transformations on the
outputs we may convert any non-local vertex into a local vertex.

For the case of two inputs and two outputs, it is
well known that the only non-trivial facets of the local polytope
$\mathcal{L}$ correspond to the CHSH inequalities \cite{fine}. There is an important connection between the
CHSH inequalities and the non-local vertices of $\mathcal{P}$.
In order to explain this, we first recall explicitly the CHSH inequalities. Let $\langle \msc{i\,j}\rangle$ be defined by
\begin{equation}\label{corr}
\langle ij \rangle=\sum_{a,b=0}^1 (-1)^{a+b}\, p_{ab|\msc{X=i,Y=j}}.
\end{equation}
Then the non-trivial facets of $\mathcal{L}$ are
equivalent to the following inequalities.
\begin{equation}\begin{split}\label{chsh}
B_{\alpha\beta\gamma} &\equiv (-1)^{\gamma}\,\langle
00 \rangle+(-1)^{\beta+\gamma}\,\langle
01\rangle \\
&+(-1)^{\alpha+\gamma}\,\langle 10
\rangle+(-1)^{\alpha+\beta+\gamma+1}\,\langle 11
\rangle \le 2,\end{split}
\end{equation}
where $\alpha$, $\beta$, $\gamma \in \{0,1\}$. For each of
the 8 Bell expressions $B_{\alpha\beta\gamma}$, the algebraic maximum is $B_{\alpha\beta\gamma}=4$.  We find that for each choice of $\alpha,\, \beta,\,\gamma$ the correlations defined by Eq.~(\ref{nonlocal}) return a value for
the corresponding Bell expression of $B_{\alpha\beta\gamma}=4$. Thus there is a one-to-one
correspondence between the non-local vertices of $\mathcal{P}$ and the
non-trivial facets of $\mathcal{L}$, with each vertex violating the corresponding CHSH inequality up to the algebraic maximum. These extremal correlations describe in a compact way the logical contradiction in the CHSH inequalities.

\subsubsection{$d$ outputs}

We now generalize the results of the preceding section. Again we have two parties, Alice and Bob, who choose from two inputs $\msc{X}$ and $\msc{Y}$ $\in \{0,1\}$ and receive outputs $a$ and $b$ with a joint probability $p_{ab|\mscs{XY}}$.
We denote the number of distinct outputs associated with inputs $\msc{X}$ and
$\msc{Y}$ by $d^A_{\mscs{X}}$ and $d_{\mscs{Y}}^B$. If Alice's input is $\msc{X}$, for example, then $a\in \{0,\ldots,d^A_{\mscs{X}}-1\}$.

\begin{theorem}\label{dimd}
The non-local vertices of $\mathcal{P}$ for two input settings
and $d_\mscs{X}^A$ and $d_\mscs{Y}^B$ outputs  are
equivalent under reversible local relabelling to
\begin{equation}\label{ddimensionalbox}
p_{ab|\mscs{XY}} = \left\{ \begin{array}{r@{\quad}c@{\quad}l} 1/k &:&
(b-a) \mathrm{\ mod\ } k = \msc{X.Y}\\
&&a,b\in\{0,\ldots,k-1\}
\\
0 &:& \mathrm{otherwise,}
\end{array} \right.
\end{equation}
for each $k\in\{2,\ldots,\min_{\mscs{X,Y}}(d_{\mscs{X}}^A,d_{\mscs{Y}}^B)\}$.
\end{theorem}

We note that the case $d_\mscs{X}^A=d_\mscs{Y}^B=2$ gives the
PR correlations we found previously.  If
$d_\mscs{X}^A=d_\mscs{Y}^B=k=d$ then the vertex violates the
$d$-dimensional generalization of the CHSH inequality
\cite{popescu+collins} up to its algebraic maximum. We call such a box a $d$-box (a more complete name would specify that the number of parties and the number of inputs per party are each two, but this simple name will do for our purposes).

{\bf Proof of Theorem~\ref{dimd}.} A probability table $p_{ab|\mscs{XY}}$ is a vertex of
$\mathcal{P}$ if and only if it is the unique solution of Eqs.~(\ref{pos}),(\ref{norm}),(\ref{nosiga}) and (\ref{nosigb}) with dim($\mathcal{P}$) of the positivity inequalities (\ref{pos})
replaced with equalities.

It will be useful to distinguish two kinds of extremal points:
partial-output vertices and full-output vertices. Partial-output
vertices are vertices for which at least one of the
$p_{a|\mscs{X}}=0$ or $p_{b|\mscs{Y}}=0$. They can be identified
with vertices of polytopes $\mathcal{P}'$ with fewer possible
outputs: ${d'}{}_{\mscs{X}}^A<d_{\mscs{X}}^A$ or
${d'}{}_{\mscs{Y}}^B<d_{\mscs{Y}}^B$. Conversely, the vertices of
a polytope $\mathcal{P}'$, with ${d'}{}_{\mscs{X}}^A<d_{\mscs{X}}^A$ or
${d'}{}_{\mscs{Y}}^B<d_{\mscs{Y}}^B$ can be extended to vertices of
$\mathcal{P}$ by mapping the outcomes of $\msc{X}'$ and
$\msc{Y}'$ to a subset of the outcomes of $\msc{X}$ and
$\msc{Y}$, and by assigning a zero probability $p_{a|\msc{X}}=0$
and $p_{b|\msc{Y}}=0$ to extra outcomes. Full-output vertices are
vertices for which all $p_{a|\mscs{X}}\neq 0$ and
$p_{b|\mscs{Y}}\neq 0$, i.e., for which all outputs contribute
non-trivially to $p_{ab|\mscs{XY}}$. Thus the extremal
points of a given two-settings polytope consist of the full-output
vertices of that polytope and, by iteration, of all the
full-output vertices of two-settings polytopes with fewer
outcomes. Hence in the following, we need construct only the full-output vertices for a polytope characterized by $d_{\mscs{X}}^A$ and $d_{\mscs{Y}}^B$.

The joint probabilities $p_{ab|\mscs{XY}}$ form a table of
$\sum_{\mscs{X,Y}}d^A_\mscs{X} d^B_\mscs{Y}$ entries. These are
not all independent because of the normalization and no-signalling
conditions. There are $4$ normalization equalities expressed by Eq.~(\ref{norm}) and $\sum_{\mscs{X}}
d_{\mscs{X}}^A+\sum_{\mscs{Y}} d_{\mscs{Y}}^B$
no-signalling equalities expressed by Eqs.~(\ref{nosiga}) and (\ref{nosigb}). But for each value of $\msc{X}$, the no-signalling condition for one of Alice's outputs can be deduced from the conditions of
normalization and no-signalling for the $d_{\mscs{X}}^A-1$
other outputs. A similar argument applies for each value of $\msc{Y}$ and Bob's outputs. Hence Eqs.~(\ref{norm}), (\ref{nosiga}) and (\ref{nosigb}) form a set of only
$4+\sum_{\mscs{X}}(d_{\mscs{X}}^A-1)+\sum_Y
(d_Y^B-1)=\sum_{\mscs{X}}(d_{\mscs{X}}^A)+\sum_Y
(d_Y^B)$ linearly independent equations. The dimension of the
no-signalling polytope is thus
\begin{equation}\label{dim2}
\mbox{dim}(\mathcal{P})=\sum_{\mscs{X,Y}=0}^1d_{\mscs{X}}^Ad_{\mscs{Y}}^B-
\sum_{\mscs{X}=0}^1d_{\mscs{X}}^A-\sum_{\mscs{Y}=0}^1d_{\mscs{Y}}^B
\ .
\end{equation}
This is the number of entries in the table $p_{ab|\mscs{XY}}$ that
must be set to zero to obtain a vertex. Moreover, to obtain a
full-output vertex, these must be chosen so that neither
$p_{a|\mscs{X}}=0$ nor $p_{b|\mscs{Y}}=0$. If we fix a particular
pair of inputs ($\msc{X},\msc{Y})$, then no more than
$d_{\mscs{X}}^Ad_{\mscs{Y}}^B-\max(d_{\mscs{X}}^A,d_{\mscs{Y}}^B)$
probabilities may be set to zero, otherwise there will be fewer
than $\max(d_{\mscs{X}}^A,d_{\mscs{Y}}^B)$ probabilities
$p_{ab|\mscs{XY}}> 0$, and thus one of Alice's or one of Bob's
outcomes will not be output for these values of $\msc{X}$ and
$\msc{Y}$. Because of the no-signalling conditions it will not be
output for the other possible pairs of inputs, so the vertex will
be a partial-output one. Overall, the maximal number of allowed
zero entries for a full-output vertex is
\begin{equation}\label{max}
Z=\sum_{\mscs{X,Y}}\left(d_{\mscs{X}}^Ad_{\mscs{Y}}^B-\max(d_{\mscs{X}}^A,
d_{\mscs{Y}}^B)\right) \ .
\end{equation}
Such a vertex is thus possible if dim$(\mathcal{P})\leq Z$, or
\begin{equation}
\sum_{\mscs{X}=0}^1d_{\mscs{X}}^A+\sum_{\mscs{Y}=0}^1d_{\mscs{Y}}^B
\geq \sum_{\mscs{X,Y}=0}^1 \max(d_{\mscs{X}}^A,d_{\mscs{Y}}^B ) \ .
\end{equation}
This condition is fulfilled (with equality) only for
$d_{\mscs{X}}^A=d_{\mscs{Y}}^B=d$, $\forall \, \msc{X,Y}$ $\in
\{0,1\}$.

We can thus restrict our analysis to $d$-outcome polytopes. The
extremal points of more general ones, where $d_{\mscs{X}}^A\neq
d_{\mscs{Y}}^B$, will be the full-output extremal points of
$d$-outcomes polytopes for
$d=2,\ldots,\min_{\mscs{X,Y}}(d_{\mscs{X}}^A,d_{\mscs{Y}}^B)$.

Using $d_{\mscs{X}}^A=d_{\mscs{Y}}^B=d$, $\forall \, \msc{X,Y}$
$\in \{0,1\}$ in the discussion before Eq.~(\ref{max}), it
follows that the dimension of a $d$-outcome polytope is $4d(d-1)$
and that for a given pair of inputs exactly $d(d-1)$ probabilities must be assigned the
value zero, or equivalently that $d$ probabilities must be $>0$.
We can therefore write the probabilities as
\begin{equation}\label{perm}
p_{ab|\mscs{XY}} \left\{ \begin{array}{r@{\quad}l}
> 0  & \mbox{if} \, b=f_{\mscs{XY}}(a)
\\
=0 & \mbox{otherwise,}
\end{array} \right.
\end{equation}
where $f_\mscs{XY}(a)$ is a permutation of the $d$ outcomes. Indeed, if
$f_\mscs{XY}(a)$ is not a permutation, then at least one of Bob's
outcomes will not be output.

We can relabel Alice's outcomes for $\msc{X}=0$ so that
$f_{01}(a)=a$, we can relabel those of Bob for $\msc{Y}=0$ so that $f_{00}(a)=a$
and finally those of Alice for $\msc{X}=1$ to have $f_{10}(a)=a$. In other
words,
\begin{equation}\label{diff0}
p_{ab|\mscs{XY}} \left\{ \begin{array}{r@{\quad}l}
> 0  & \mbox{if} \, (b-a)\mathrm{\ mod\ }d = 0
\\
=0 & \mbox{otherwise,}
\end{array} \right.
\end{equation}
for $(\msc{X,Y}) \in \{ (0,0),(0,1),(1,0)\}$. It remains to determine $f_{11}$. It must be chosen so that the probability table $p_{ab|\mscs{XY}}$ is uniquely determined, i.e.,
so that specific values are assigned to the probabilities
different from zero. In fact, it is easy to show that this can only be the case if the
permutation $f_{11}$ is of order $d$, i.e., $f_{11}^k(a)=a$ only for $k=0 \mathrm{\ mod\ }d$.

The only remaining freedom in the relabelling of the outcomes so
that property (\ref{diff0}) is conserved, is to relabel simultaneously the outputs for all four possible inputs. We can relabel them globally so
that $f_{11}(a)=(a+1) \mathrm{\ mod\ }d$. This implies that $p_{ab|11}=1/d$
if $(b-a)\mathrm{\ mod\ }d=1$. This completes the proof. \hfill $\Box$

\subsection{Resource conversions}\label{resconvtwosett}

In the preceding section we found all the vertices of the
no-signalling polytope for bipartite, two-input boxes. As described in the introduction, the ethos adopted in this work is that boxes (in particular, non-local boxes) can be regarded as an information theoretic resource, and investigated as such. Useful comparisons can be drawn with other information theoretic resources, including shared random data \cite{ieee2}, shared secret data \cite{ieee1,collinspopescu}, and entanglement \cite{nielsenchuang}. In each case, there is a convex set of possible states and a notion of interconversion between different states. There is also a notion of interconversion between different resources. Each resource is useful for some task(s) and can be quantified via some measure(s). Some of this is illustrated in Table~\ref{table}.
\begin{table}
\begin{tabular}{l|l|l}
\multicolumn{1}{c|}{Resource}&\multicolumn{1}{c|}{Instantiation}&\multicolumn{1}{c}{Quantitative measure} \\ \hline 
Shared random data & Random variables & Mutual information \\
Shared secret data & Random variables & Secrecy rate \\
Entanglement & Quantum states & Entanglement cost \\
Non-locality & Boxes & Classical simulation cost
\end{tabular}
\caption{Comparison of information theoretic resources.\label{table}}
\end{table}
Note that the quantitative measures given are not the only possibilities. Note also that even if the given measure vanishes, a useful resource may still be present. Thus uncorrelated random variables can still be useful (as local randomness), as can separable quantum states (for various things), as can local boxes (as local or shared randomness).

In light of this, it is natural to ask, what interconversions between boxes are possible, and what would be a good measure of the non-locality of a box? To the second question, several answers suggest themselves, such as the amount of classical communication needed to simulate the box (given that the only other resource is shared random data), and the degree of violation of Bell inequalities \cite{bellcomm}. In this work, however, we concentrate on the first question - partly because it is independently interesting, and partly because an understanding of possible interconversions is a prerequisite for a good understanding of quantitative measures. 

The problem that we consider, then, is whether one can simulate one type of box, using one or more copies of another type as a resource. Local operations such as relabelling are of course allowed. As non-locality is the resource that we have in mind, it is also natural to allow the parties free access to local boxes (i.e., to local and shared randomness). We note, however, that neither local nor shared randomness can help if the box to be simulated is a vertex \footnote{This is easy to see. For each value of the local or shared randomness, one can write down the box that is simulated, conditioned on that value occurring. The box simulated by the overall protocol is then the average of these conditional boxes, with the average taken over the possible values of the randomness. But if this box is a vertex, then each of the conditional boxes must be the same vertex, and the protocol could have been carried out without the randomness.}, thus none of the protocols we describe below make use of this. We make the assumption that communication between the parties is not allowed.

\begin{figure}
\scalebox{1}{\includegraphics{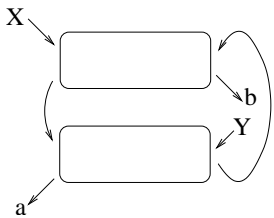}}
\caption[]{An example of how two parties that are given two boxes may process locally their inputs and outputs. They result in simulating another type of box with inputs $\msc{X,Y}$ and outcomes $a,b$. Note that due to the no-signalling condition, the parties can use their two boxes with a different time ordering. \label{twoboxes}}
\end{figure}

In general, outputs for one box can be used as inputs for another box. This allows non-trivial protocols to be constructed. As an interesting logical possibility, we note that the temporal order in which each party uses the boxes need not be the same, and that this allows loops to be constructed that would be ill-defined if it were not for the no-signalling condition. (Thus if signalling boxes were to be considered, our stipulation that outputs are obtained immediately after inputs would have to be altered.) Such a loop is illustrated in Fig.~\ref{twoboxes}.
In all of the protocols presented below, however, the parties use the boxes in the same temporal order. 

In the following, we will describe three simple examples. We show that given a $d$-box and a $d'$-box, we can simulate a $dd'$-box. We will also show that given a $dd'$-box, we can simulate one $d$-box. Finally, an unlimited supply of $d$-boxes can simulate a $d'$-box to arbitrarily high precision. 
In addition, we will describe a negative result: it is not in general possible to go \emph{reversibly} from $n$ $d$-boxes to $m$ $d'$-boxes, where $d\neq d'$. Although we only prove this for exact transformations, we believe a similar result should hold even if transformations need only be exact in an asymptotic limit. It follows from this that $d$ and $d'$-boxes are ultimately inequivalent resources and that in our context, it is inappropriate to suppose that they can be characterized by a single numerical measure of non-locality \footnote{Similar considerations apply to the other resources we have mentioned. In the case of entanglement, for example, reversible interconversions are not in general possible for mixed states, thus there is no unique measure of entanglement for mixed states. In the case of shared random data, interconversions by local operations are rather limited and provide no very meaningful measure of shared randomness. However, if one expands the set of operations that Alice and Bob are allowed, then the picture changes. Thus in the case of shared random data, allowing that Alice and Bob can communicate classically, while demanding that the communication must be subtracted at the end, gives an operational meaning to the mutual information \cite{ieee2}. Inspired by this, it may be interesting to consider conversions between boxes, with classical communication allowed but subtracted at the end, or indeed conversions between entangled quantum states with quantum communication allowed but subtracted at the end. We do not pursue these questions here.}.  

\begin{figure}
\scalebox{1}{\includegraphics{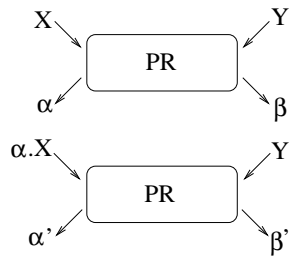}}
\caption[]{Making a $4$-box from two PR boxes. Alice inputs $\msc{X}$ into the first box and $\alpha.\msc{X}$ into the second, while Bob
inputs $\msc{Y}$ into both boxes. Alice's output is given
by $a=2\alpha' + \alpha$ and Bob's by $b=2\beta' + \beta$.\label{chshboxesfig}}
\end{figure}

Suppose first, then, that Alice and Bob have one $d$-box and one $d'$-box and they wish
to simulate one $dd'$-box. Simulate means that for each value of $\msc{X}\in\{0,1\}$, a procedure should be defined for Alice, using the $d$ and $d'$-boxes, that eventually enables her to determine the value of an output $a\in\{0,\ldots,dd'-1\}$. Similarly for Bob; for each value of $\msc{Y}$ there is an eventual output $b$. The joint probabilities for $a$ and $b$ should satisfy Eq.~(\ref{ddimensionalbox}) (with $dd'$ inserted instead of $d$ where necessary).\\[1em]
\mbox{{\bf Protocol 1: 1 $d$-box and 1 $d'$-box $\rightarrow$ 1 $dd'$-box}}\\*
\emph{Alice.} Alice inputs $\msc{X}$ into the $d$-box, obtaining outcome $\alpha$. She then inputs $\msc{X}$ into the $d'$-box if $\alpha=d-1$, and inputs $0$ into the $d'$-box otherwise, obtaining an output $\alpha'$. Alice's output for the protocol is $a=\alpha' d + \alpha$.\\
\emph{Bob.} Bob inputs $\msc{Y}$ into the $d$-box, obtaining output $\beta$, and inputs $\msc{Y}$ into the $d'$-box, obtaining output $\beta'$. His output for the protocol is then $b=\beta' d+\beta$.\\
\\
Protocol 1 is illustrated in Fig.~\ref{chshboxesfig} for the case $d=d'=2$. 
We indicate briefly why this protocol works. Recall that a $dd'$-box satisfies $(b-a)\mod dd'=\msc{XY}$. Write $a=\alpha' d+\alpha$ and $b=\beta' d+\beta$, where $\alpha$ can take values $\alpha=0,\ldots,d-1$, and $\alpha'$ can take values $\alpha'=0,\ldots,d'-1$, and so on. We see that the condition satisfied by a $dd'$-box is equivalent to
\begin{eqnarray}\label{dd}
\beta -\alpha\mod d &=& \msc{XY} \nonumber \\
\beta'-\alpha'\mod d'&=&\left\{ \begin{array}{r@{\quad}c@{\quad}l} \msc{XY} &:& \alpha=d-1\\
0 &:& \mathrm{otherwise.}\end{array} \right.
\end{eqnarray}
Protocol 1 is designed precisely to satisfy this condition. It is then not difficult to check that the correct probabilities are reproduced. 

We note next that it is easy to convert one $dd'$-box into one $d$-box. \\[1em] 
{\bf Protocol 2: 1 $dd'$-box $\rightarrow$ 1 $d$-box}\\*
\emph{Alice.} Alice inputs $\msc{X}$ into the $dd'$-box, obtaining an output $\alpha$. Her output for the protocol is then $a=\alpha\mod d$.\\
\emph{Bob.} Bob inputs $\msc{Y}$ into the $dd'$-box, obtaining an output $\beta$. His output for the protocol is $b=\beta\mod d$. \\
\\
Again, it is not difficult to check that $(b-a)\mod d=\msc{XY}$, and that the correct probabilities are reproduced.

Now we show how $n$ $d$-boxes can be used to simulate a $d'$-box to arbitrarily high precision. This is done using a combination of Protocols 1 and 2. \\[1em] 
{\bf Protocol 3: $n$ $d$-boxes $\rightsquigarrow$ 1 $d'$-box}\\*
Alice and Bob begin by using the $n$ $d$-boxes to simulate a $d^n$-box, as per Protocol 1. Call the outputs for the $d^n$-box $\alpha$ and $\beta$. They satisfy $(\beta-\alpha)\mod d^n=\msc{XY}$. Alice and Bob now use Protocol 2 to obtain something close to a $d'$-box: the final outputs are $a=\alpha\mod d'$ and $b=\beta\mod d'$. \\
\\
If $d^n=kd'$ for some positive integer $k$, this protocol works exactly. Otherwise, one can calculate the errors resulting in Protocol 3. Denote by $k$ the largest integer such that $kd'\leq d^n$. Now we have that if $\msc{X}=0$ or $\msc{Y}=0$, then $(b-a)\mod d'=0$ as required. However, the probabilities are skewed by an amount $\propto 1/k \approx d'/d^n$. If $\msc{X}=\msc{Y}=1$, then the probabilities are skewed in a similar manner. But in addition we have that if $b=d^n-1$, then $(b-a)\mod d'=1$ is not satisfied with probability $1/d^n$. The important thing here is that all errors tend to zero exponentially fast as $n$ becomes large.

We have seen several examples of how interconversions between non-local extremal boxes are possible using only local operations. It is also interesting to consider how boxes may be simulated using only classical communication (CC) and shared random data (SR), i.e., without other boxes. For example, we can see that one $d$-box may be simulated with one bit of 1-way communication and $\log_2 d$ bits of shared randomness. \\[1em] 
{\bf Protocol 4: 1 bit CC and $\log_2 d$ bits SR $\rightarrow$ 1 $d$-box}\\*
Alice and Bob share a random variable $\alpha \,\in\{0,\cdots,d-1\}$, where $\alpha$ takes all its possible values with equal probability $1/d$.\\
\emph{Alice.} Alice sends her input $\msc{X}$ to Bob and outputs $a=\alpha$.\\
\emph{Bob.} Bob, knowing $\msc{X}$ and $\alpha$, outputs $b=(\alpha +\msc{X.Y}) \mod d$.\\
\\
This protocol is optimal regarding the amount of 1-way communication exchanged. This is a consequence of the following lemma, which places a lower bound on the amount of communication needed to simulate boxes. The lemma is used in the proof of Theorem~\ref{transformationtheorem}, our final main result for this section.
\begin{lemma}\label{lemma0}
The simulation of $n$ $d-$boxes using 1-way communication requires at least $n$ bits of communication if shared randomness is available, and $n+n\log_2 d$ bits without shared randomness.
\end{lemma}
\noindent
{\bf Proof.} Note that this bound can be achieved using Protocol 4 for each of the $n$ boxes, replacing if necessary $n\log_2 d$ bits of shared randomness by $n\log_2 d$ bits of communication from Alice to Bob.

Let us show that this amount of communication is necessary. Suppose first that both parties have access to shared random data and that communication is allowed from Alice to Bob. Bob's output is thus $b=b(Y,C,r)$ where $Y=\msc{Y}_1\ldots\msc{Y}_n$ are the joint inputs for Bob, $C$ is the communication and $r$ the shared data. Note simply that for Alice, there are $2^n$ possible joint inputs into $n$ $d$-boxes. If Alice is sending fewer than $n$ bits, there will be at least one pair of joint inputs for which her communication is the same. Call them $X_1$ and $X_2$. A careful examination of the definition of a $d$-box reveals that there will be at least one joint input of Bob's into the $n$ boxes such that his output must be different according to whether Alice's input was $X_1$ or $X_2$. Thus $<n$ bits of communication are not sufficient.

If Alice and Bob do not have access to shared randomness, then Bob's output is of the form $b=b(Y,C)$. The proof then follows by an argument similar to the one used above, noting that for Alice there are $2^{n+n\log_2 d}$ possible joint input-output pairs $(X,A)$. \hfill$\square$

These types of considerations will help us to establish the final result of this section.
\begin{theorem}\label{transformationtheorem}
It is in general impossible, using local reversible operations, exactly to transform $n$ $d$-boxes into $m$ $d'$-boxes.
\end{theorem}

The theorem follows from the following two lemmas.
\begin{lemma}\label{lemma2}
Using $n$ $d$-boxes, Alice and Bob can exactly simulate at most $n$ $d'$-boxes, for $d\geq d'$.
\end{lemma}
\begin{lemma}\label{lemma3}
Using $n$ $d'$-boxes, Alice and Bob can exactly simulate at most $n(1+\log_2 d')/(1+\log_2 d)<n$ $d$-boxes for $d'\leq d$.
\end{lemma}
\noindent
{\bf Proof.} We prove Lemma~\ref{lemma2} as follows. We know that we can simulate $n$ $d$-boxes with $n$ bits of communication and $n\log d$ bits of shared randomness. Suppose that there were a protocol using only local operations that could convert $n$ $d$-boxes into $N$ $d'$ boxes, for some $d'\leq d$, where $N>n$. Then, by combining the simulation of the $d$-boxes with the protocol for their conversion, we would have constructed a protocol for simulating $N$ $d'$-boxes using only $n$ bits of communication, in contradiction with Lemma~\ref{lemma0}. The proof of Lemma~\ref{lemma3} is very similar. Note that we can simulate $n$ $d'$-boxes with $n+n\log_2 d'$ bits of classical communication and no shared randomness. Suppose that there were a protocol that converts $n$ $d'$-boxes into $N$ $d$-boxes, for some $d\geq d'$, where 
$N>n(1+\log_2 d')/(1+\log_2 d)$. As argued above, it follows from the fact that $d$-boxes are vertices that this protocol would not need any additional shared randomness. Then we would have constructed a protocol for simulating $N$ $d$-boxes using only $n+n\log_2 d'$ bits of communication and no shared randomness, again in contradiction with Lemma \ref{lemma0}. \hfill $\square$

\section{Three party correlations}\label{threepartpoly}

\subsection{Definitions}

In this section, we generalize the considerations of the previous sections to consider tripartite correlations. As before, we consider that correlations are produced by a black box with specified inputs and outputs, but now the box is assumed to be shared between three separated parties, $A$, $B$ and $C$. 
\paragraph*{The no-signalling polytope.} A box is defined by joint probability distributions $p_{abc|XYZ}$, which satisfy positivity,
\begin{equation}\label{pos3}
p_{abc|\mscs{XYZ}}\ge 0 \,\qquad \forall \,a,b,c,\msc{X,Y,Z}
\end{equation}
normalization,
\begin{equation}
\label{norm3} \sum_{a,b,c} p_{abc|\mscs{XYZ}}=1 \qquad \forall \
\msc{X,Y,Z}
\end{equation}
and no-signalling. With three parties it is possible to imagine
various types of communication, and correspondingly there are
different types of no-signalling conditions. Obviously we require that $A$ cannot signal to $B$ or $C$ (and cyclic permutations). We should also, however, require the stronger condition that if the systems $B$ and $C$ are combined, then $A$ cannot signal to the resulting composite system $BC$. This is expressed by
\begin{equation}\label{nosig3}
\sum_{a}p_{abc|\mscs{X,Y,Z}} =\sum_{a}p_{abc|\mscs{X',Y,Z}} \quad \forall\, b,c,\msc{Y,Z,X,X'},
\end{equation}
where, again, we include cyclic permutations.
Finally, note that if systems $A$ and $B$ are combined, the resulting composite system $AB$ should not be able to signal to $C$. This type of condition does not require a separate statement, however, as it already follows from Eq.~(\ref{nosig3}). Indeed, using the fact that $A$ cannot signal to $BC$ and that $B$ cannot signal to $AC$, we deduce
\begin{eqnarray}
\sum_{a,b}p_{abc|\mscs{X,Y,Z}} &=& \sum_{a,b}p_{abc|\mscs{X',Y,Z}} \quad \forall\, b, c,\msc{X,X',Y,Z}\nonumber \\
&=&\sum_{a,b}p_{abc|\mscs{X',Y',Z}} \quad \forall\, c,\msc{X,X',Y,Y',Z}, \nonumber \\
\end{eqnarray}
which is the condition that $AB$ cannot signal to $C$.
Hence the only conditions we need to impose on a tripartite box are those of Eqs.~(\ref{pos3}), (\ref{norm3}) and (\ref{nosig3}). The set of boxes satisfying these conditions is the polytope $\mathcal{P}$. 

\paragraph*{Locality conditions.}
In the tripartite case, as well as different types of no-signalling condition, there are different types of locality condition. First, a box is fully local if the probabilities can be written in the form
\begin{equation}
p_{abc|\mscs{XYZ}}=\sum_{\lambda} p_{\lambda}\,p_{a|X}(\lambda)\,p_{b|Y}(\lambda)\,p_{c|Z}(\lambda).
\end{equation}
The set of such boxes is a convex polytope denoted $\mathcal{L}$. Second, we say that a box is two-way local if either there exists a bi-partition of the parties, say $AB$ versus $C$, such that the composite system $AB$ is local versus $C$, or if the box can be written as a convex combination of such boxes, i.e.,
\begin{eqnarray}\label{twnl}
p_{abc|\mscs{XYZ}}&=& p_{12} \, \sum_{\lambda_{12}}\,p_{\lambda_{12}}\,
p_{ab|\mscs{XY}}(\lambda_{12})\,p_{c|\mscs{Z}}(\lambda_{12})\nonumber\\ 
&+& p_{13} \, \sum_{\lambda_{13}}\,p_{\lambda_{13}}\,p_{ac|\mscs{XZ}}(\lambda_{13}) \, p_{b|\mscs{Y}}(\lambda_{13})\nonumber\\
&+& p_{23}\,\sum_{\lambda_{23}}\,p_{\lambda_{23}}\,p_{bc|\mscs{YZ}}(\lambda_{23}) \, p_{a|\mscs{X}}(\lambda_{23}),
\end{eqnarray}
where $p_{12}+p_{23}+p_{13}=1$. The set of such boxes is again a convex polytope, denoted $\mathcal{L}2$. Finally, any box that cannot be written in this form demonstrates genuine three-way non-locality. We have that $\mathcal{L}\subset\mathcal{L}2\subset\mathcal{P}$ and also that $\mathcal{L}\subset\mathcal{Q}\subset\mathcal{P}$.

In the following, we restrict our attention to the case $a,b,c,\mscs{X,Y,Z} \in
\{0,1\}$. We find the vertices of the polytope $\mathcal{P}$ and point out some connections with three-party Bell-type inequalities. Finally we consider some examples of interconversions, in particular of how to construct tripartite boxes using PR boxes as a resource.

\subsection{Two inputs and two outputs}
For the tripartite boxes with two inputs and two outputs per observer, Eq.~(\ref{norm3}) expresses 8 normalization constraints, and Eq.~(\ref{nosig3}) expresses $3\times12=48$ no-signalling constraints. However, as in the bipartite case, there is also some further redundancy; there turn out to be 38 independent constraints. Therefore the dimension of this polytope is dim $\mathcal{P}=2^6-38=26$.

Finding the vertices of a polytope given its facets is the so called ``vertex enumeration problem" for which several algorithms are available, although they are efficient only for low dimensional problems. 
We determined the extreme points of our three-party polytope, both with {\emph{Porta} \cite{porta} and {\emph{cdd} \cite{cdd}. It turns out that there are 46 classes of vertices, where vertices within one class are equivalent under local relabelling operations and permutations of the parties. These 46 classes of extreme points can be divided into three categories: local, two-way local and three-way non-local. 

\paragraph*{Local vertices.} This category contains boxes for which $A$'s, $B$'s and $C$'s outputs are all deterministic. They all belong to the same class under reversible local operations, a representative of which is:
\begin{equation}\label{locclass}
p_{abc|\mscs{XYZ}}=\left\{ \begin{array}{r@{\quad}c@{\quad}l} 1 &:& a=0,\, b=0,\, c=0\\
0 &:& \mathrm{otherwise.}\end{array} \right.
\end{equation}

\paragraph*{Two-way local vertices.} In view of the preceding discussion for bipartite correlations, there is only one class of extremal two-way local correlations that are not fully local. This is because if a box is a vertex, there can be only one term in the decomposition on the right hand side of Eq.~(\ref{twnl}). Then it follows from Theorem~\ref{dimd} that this term must describe a PR box shared between two parties, along with a deterministic outcome for the third party. Thus any box of this type is equivalent under local relabellings and permutations of parties to
\begin{equation}\label{twclass}
p_{abc|\mscs{XYZ}} = \left\{ \begin{array}{r@{\quad: \quad}l} 1/2
& a\oplus b=\msc{X.Y} \mathrm{\ and\ } c=0\\
0 & \mathrm{otherwise.}
\end{array} \right.
\end{equation}

\paragraph*{Three-way non-local vertices.} This category contains genuine three-party non-local extremal correlations. It is much more complex than the two above, since it comprises 44 different classes of vertices. Out of these, we mention 3 classes of particular interest. The first class can be expressed as
\begin{equation}\label{dprclass}
p_{abc|\mscs{XYZ}} = \left\{ \begin{array}{r@{\quad}c@{\quad}l} 1/4
&:& a\oplus b\oplus c\\ && \quad =\msc{X.Y}\oplus\msc{X.Z}\\
0 &:& \mathrm{otherwise.}
\end{array} \right.
\end{equation}
If we imagine that $B$ and $C$
form a composite system with input $\msc{Y}\oplus\msc{Z}$ and output
$b\oplus c$, then this is a PR box shared between $A$ and $BC$. We refer to them as ``X(Y+Z)'' boxes. 

Correlations in the second class are equivalent to
\begin{equation}\label{svetlichnyclass}
p_{abc|\mscs{XYZ}} = \left\{ \begin{array}{r@{\quad}c@{\quad}l} 1/4
&:& a\oplus b\oplus c\\ && \quad =\msc{X.Y}\oplus \msc{Y.Z}\oplus \msc{X.Z}\\
0 &:& \mathrm{otherwise.}
\end{array} \right.
\end{equation}
We call them ``Svetlichny'' correlations (for reasons explained below).

Finally, the third class contains what we call ``XYZ'' correlations.
\begin{equation}\label{merminclass}
p_{abc|\mscs{XYZ}} = \left\{ \begin{array}{r@{\quad: \quad}l} 1/4
& a\oplus b\oplus c=\msc{X.Y.Z}\\
0 & \mathrm{otherwise.}
\end{array} \right.
\end{equation}
The XYZ correlations are special because, as W.~van~Dam pointed out to us \cite{vandamprivcomm}, they can be used to solve any three party communication complexity problem with only 1 bit broadcast by each party. He also pointed out that they have a natural generalization to $n$ parties: $a_1\oplus a_2\oplus\cdots\oplus a_n=\msc{X_1}.\msc{X_2}\ldots\msc{X_n}$, where $\msc{X_i}\in\{0,1\}$ is the input of party $i$ and $a_i\in\{0,1\}$ the output of party $i$. These $n$-party correlations can be used to solve any $n$ party communication complexity problem with 1 bit broadcast by each party. They can be constructed from a supply of PR boxes.

We conclude this section with some remarks on these correlation vertices and known multipartite Bell-type inequalities. First, each of the X(Y+Z), XYZ, and Svetlichny boxes violates the Mermin-Klyshko inequality \cite{mermin,klyshko} up to the algebraic maximum. Second, we recall that inequalities can be written down that detect genuine three-way non-locality. One such is the Svetlichny inequality \cite{Svetlichny}.
If we define $\langle\msc{i\,j\,k}\rangle$ by
\begin{equation}
\langle ijk \rangle=\sum_{a,b,c} (-1)^{a+b+c}\,
p_{a,b,c|\mscs{X=i,Y=j,Z=k}},
\end{equation}
then the Svetlichny inequality is
\begin{equation}\begin{split}
M=-\langle 0&00 \rangle+\langle 001 \rangle+\langle 001
\rangle+\langle 011 \rangle \\ & +\langle 100 \rangle+\langle 101
\rangle+\langle 110 \rangle-\langle 111 \rangle \leq 4.
\end{split}\end{equation}
Any local or two-way local box must satisfy this inequality. Quantum mechanically we can obtain $M=4\,\sqrt{2}$ using a Greenberger-Horne-Zeilinger (GHZ) state \cite{ghz} (although note that different measurements are needed from those that produce the well known GHZ paradox \cite{roberts}). X(Y+Z) boxes do not violate the Svetlichny inequality (although they must violate some Svetlichny-type inequality as they are three-way non-local). Svetlichny boxes give $M=8$, the algebraic maximum of the expression (hence their name); XYZ correlations give $M=6$. 

From the fact that some quantum states violate the Svetlichny inequality, we can conclude that in the two-input two-output case, $\mathcal{Q}\nsubseteq{L}2$. From the fact that bipartite correlations can be more non-local than quantum mechanics allows, we can also conclude that $\mathcal{L}2\nsubseteq\mathcal{Q}$.  

\subsection{Simulating tripartite boxes}

We consider how we may simulate some of these tripartite boxes, using a supply of PR boxes as a resource. We will give three examples, showing how to simulate an X(Y+Z) box with two PR boxes, a Svetlichny box with three PR boxes, and an XYZ box with three PR boxes.

First, suppose that two PR boxes are shared, with box~1 between Alice and Bob and box~2 between Alice and Charles. The following protocol shows how the three observers may simulate one X(Y+Z) box (see Fig.~\ref{2pr}).
\begin{figure}
\includegraphics{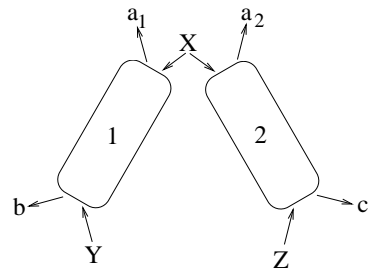}
\caption{Making an X(Y+Z) box from 2 PR boxes. Alice outputs $a=a_1\oplus a_2$, Bob outputs $b$ and Charles outputs $c$.\label{2pr}}
\end{figure}
\\[1em]
{\bf Protocol 5: 2 PR boxes $\rightarrow$ 1 X(Y+Z) box}\\
\emph{Alice.} Alice inputs $\msc{X}$ into box 1 and box 2, obtaining outputs $a_1$ and $a_2$. She then outputs $a=a_1\oplus a_2$.\\
\emph{Bob.} Bob inputs $\msc{Y}$ into box 1, obtaining output $b$.\\
\emph{Charles.} Charles inputs $\msc{Z}$ into box 2 obtaining output $c$. \\

The protocol works because
\begin{equation}
a\oplus b\oplus c=a_1\oplus a_2\oplus b\oplus c=\msc{X.Y}\oplus \msc{X.Z}.
\end{equation}

Suppose now that three PR boxes are shared, with box~1 between Alice and Bob, box~2 between Alice and Charles, and box~3 between Bob and Charles. Protocol~6 (summarized in Fig.~\ref{svetlichnyfig}) allows them to simulate one Svetlichny box.
\begin{figure}
\includegraphics{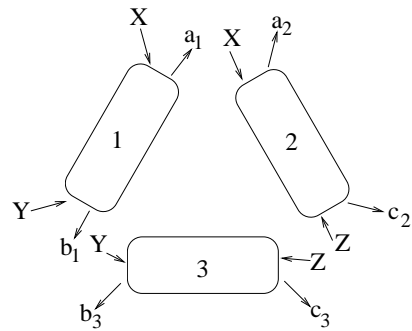}
\caption{Making a Svetlichny box from 3 PR boxes. Alice outputs $a=a_1\oplus a_2$, Bob outputs $b=b_1\oplus b_3$ and Charles outputs $c=c_2\oplus c_3$.\label{svetlichnyfig}}
\end{figure}
\\[1em]
{\bf Protocol 6: 3 PR boxes $\rightarrow$ 1 Svetlichny box}\\
\emph{Alice.} Alice inputs $\msc{X}$ into both box 1 and box 2, obtaining $a_1$ and $a_2$. Her final output is $a=a_1\oplus a_2$.
\emph{Bob.} Bob inputs $\msc{Y}$ into both box 1 and box 3, obtaining $b_1$ and $b_3$. His final output is $b=b_1\oplus b_3$.\\
\emph{Charles.} Charles inputs $\msc{Z}$ into both box 2 and box 3, obtaining $c_2$ and $c_3$. His final output is $c=c_2\oplus c_3$. \\

This works because
\begin{eqnarray}
a\oplus b\oplus c&=&a_1\oplus b_1\oplus b_3\oplus c_3\oplus a_2\oplus c_2\nonumber\\
&=&\msc{X.Y}\oplus \msc{Y.Z}\oplus \msc{X.Z}.
\end{eqnarray}

Protocol~7 (summarized in Fig.~\ref{merminfig}) shows how to simulate one XYZ box using three PR boxes.
\begin{figure}
\includegraphics{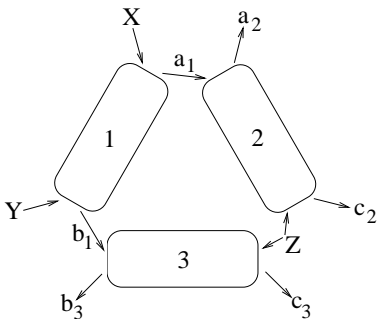}
\caption{Making an XYZ box from 3 PR boxes. Alice outputs $a=a_2$, Bob outputs $b=b_3$ and Charles outputs $c=c_2\oplus c_3$.\label{merminfig}}
\end{figure}
\\[1em]
{\bf Protocol 7: 3 PR boxes $\rightarrow$ 1 XYZ box}\\
\emph{Alice.} Alice inputs $\msc{X}$ into box 1, obtaining an output $a_1$. She then inputs $a_1$ into box 2, obtaining output $a_2$. Alice's output for the protocol is $a=a_2$.\\
\emph{Bob.} Bob inputs $\msc{Y}$ into box 1, obtaining an output $b_1$. He then inputs $b_1$ into box 3, obtaining output $b_3$. Bob's output for the protocol is $b=b_3$.\\
\emph{Charles.} Charles inputs $\msc{Z}$ into both boxes 2 and 3, obtaining outputs $c_2$ and $c_3$. Charles' output for the protocol is $c=c_2\oplus c_3$.\\

The protocol works because
\begin{equation}
a\oplus b\oplus c=a_2\oplus b_3\oplus c_2\oplus c_3=\msc{Z}.a_1\oplus \msc{Z}.b_1=\msc{X.Y.Z}.
\end{equation}

Finally, we note that it is of course possible to perform conversions among tripartite boxes. For example, it is easy to see how to make one Svetlichny box using two XYZ boxes. The protocol is obvious once it is realized that a Svetlichny box is locally equivalent to a box defined by Eq.~(\ref{svetlichnyclass}) with $\msc{XY}\oplus \msc{YZ}\oplus\msc{XZ}$ on the right hand side replaced by $\msc{XYZ}\oplus (1\oplus\msc{X})(1\oplus\msc{Y})(1\oplus\msc{Z})$. We omit the details.  

\subsection{Non-locality and the environment}\label{env}
Suppose that we have some three party no-signalling
distribution $p_{abe|\mscs{XYE}}$ with parties $A,B$ and $E$. We will show that if the reduced probability distribution $p_{ab|\mscs{XY}}=\sum_e p_{abe|\mscs{XYE}}$ is a vertex of the bipartite no-signalling polytope, then the composite system $AB$ is local versus $E$. This is analogous to the result that pure quantum states cannot be entangled with a third party or the environment. It means that extremal non-local correlations cannot be correlated to any other system. (Note that this raises interesting new possibilities for cryptography. These are investigated in Ref.~\cite{qkd}.)
 
By Bayes' theorem
\begin{eqnarray}
p_{abe|\mscs{XYE}}&=&p_{ab|\mscs{XYEe}}\,p_{e|\mscs{XYE}}\nonumber \\
&=&p_{ab|\mscs{XYEe}}\,p_{e|\mscs{E}}
\end{eqnarray}
where we have used the fact that $AB$ cannot signal to $E$ to deduce the second equality.
The condition that $E$ cannot signal to $AB$ implies
\begin{eqnarray}
p_{ab|\mscs{XY}}&=&\sum_e p_{abe|\mscs{XYE}} \quad \qquad \forall E \nonumber \\
&=&\sum_e p_{ab|\mscs{XYE}e}\,p_{e|\mscs{E}} \quad \forall E
\end{eqnarray}
For each value $E$, the last equality provides a convex decomposition of $p_{ab|\mscs{XY}}$ in terms of non-signalling correlations, with $e$ playing the role of the shared randomness. Since we supposed that $p_{ab|\mscs{XY}}$ is extremal, this decomposition is unique and $p_{ab|\mscs{XYE}e}=p_{ab|\mscs{XY}}$ $\forall e,E$. We then deduce
\begin{equation}
p_{abe|\mscs{XYE}}=p_{ab|\mscs{XY}}\, p_{e|E},
\end{equation}
i.e., that $AB$ is uncorrelated with $E$.

A natural question that we leave as an open problem is whether the converse is true: if $p_{ab|\mscs{XY}}$ is in the interior of the no-signalling polytope, is it always possible to extend it to a tripartite distribution $p_{abe|\msc{XYE}}$ such that $AB$ is non-local versus $E$? (It is always possible, if $p_{ab|\msc{XY}}$ is not a vertex, to write it as $p_{ab|\msc{XY}}=\sum_ep_{abe|\msc{XYE}}$, where $E$ takes the single value $E=0$. One can also require that $E$ take several values, in such a way that $p_{abe|\msc{XYE}}$ is non-signalling. What is non-trivial is the requirement that $p_{abe|\msc{XYE}}$ is non-local in the partition $AB$ versus $E$. We do not know if this is possible in general.)

\section{Discussion and open questions}\label{opquest}

In conclusion, we have defined non-signalling correlation boxes and investigated their potential as an information theoretic resource. Once the structure of the set of such boxes is understood as a convex polytope, it is clear that there are analogies with other information theoretic resources, in particular the resource of shared quantum states (with non-locality taking the place of entanglement). With this in mind, we have shown how various interconversions between boxes are possible. The set of multipartite boxes in particular appears very rich. Finally, we furthered the analogy with quantum states by demonstrating how non-locality is monogamous, in much the same way that entanglement is monogamous. We finish with some open questions.   

\paragraph*{Non-local vertices and Bell inequalities.}
We saw in Sec.~\ref{twotwotwo} that for the two-settings two-outcomes polytope there is a one-to-one correspondence between extremal non-local correlations and facet Bell inequalities (non-trivial facets of the local polytope). One might wonder whether this one-to-one correspondence holds in general. It appears, however, that for more complicated situations, involving more possible inputs or outcomes, it does not. It would be interesting to investigate what is the precise relation between non-local vertices and facet Bell inequalities. This might help understand further the geometrical structure of non-local correlations.

\paragraph*{Other vertices.}
We have given a complete characterization of two-inputs extremal non-local boxes in the bipartite case and presented some examples in the tripartite case. In general, one might also consider extremal boxes involving more inputs, more outcomes or more parties. 

For instance, a natural way to generate more complex boxes is by taking products of simpler ones. Suppose Alice and Bob have access to two boxes $p^0_{a_0b_0|\mscs{X}_0\mscs{Y}_0}$ and $p^1_{a_1b_1|\mscs{X}_1\mscs{Y}_1}$, where for simplicity we consider that there are $M$ possible inputs and $d$ possible outputs for each box. If Alice inputs $\msc{X}_0$ and $\msc{X}_1$ in each of the two boxes and outputs $a=d\,a_1+a_0$ and similarly for Bob, they have now produced a non-local box with $M^2$ inputs and $d^2$ outputs $p_{ab|\mscs{XY}}= p^0_{a_0b_0|\mscs{X}_0\mscs{Y}_0}\, . \, p^1_{a_1b_1|\mscs{X}_1\mscs{Y}_1}$, where $\msc{X}=M\,\msc{X}_1+\msc{X}_0$ and similarly for $\msc{Y}$. If the two original boxes were extremal for the $(M,d)$ polytope will the product be extremal for the $(M^2,d^2)$ polytope? In the case of quantum states, the analogous result of course holds - a product of two pure states is itself a pure state. We have been able to show that in the case of boxes, the result holds provided that we restrict to extremal boxes with the following property: the output of one party is uniquely determined when the two inputs and the other party's output are specified. This is true for all the vertices presented in this paper. Plausibly it is true for all vertices, but this is not proven. 

\paragraph*{Interconversions.}
We have so far been able to achieve only a limited set of
interconversions between extremal boxes. This is especially true for the three party case, where there are 46 classes of vertices and we have investigated only 5 of these. 
Understanding what kinds of interconversions between extremal boxes are possible is necessary to appraise their relative power as an information-theoretic resource. 

The motivation is also to answer the general question of whether there exist inequivalent types of non-local correlations. Note for instance that the three-way non-local correlations of Eqs.~(\ref{dprclass}), (\ref{merminclass}) and (\ref{svetlichnyclass}) cannot be reduced to two-way non-local ones using only local operations. This follows from the fact that the outcomes for two out of the three parties are totally independent of one another (unless the outcome of the third party is communicated to them). In this sense genuinely tripartite extremal correlations and bipartite extremal correlations belong to inequivalent classes. Are there inequivalent classes of bipartite extremal correlations? In other words, are there two bipartite extremal boxes, such that one cannot simulate the other even approximately, no matter how many copies are available?   

Another problem is whether all bipartite and multipartite correlations can be constructed using PR boxes, as is the case for all the extremal boxes presented in this paper (and thus also for probabilistic mixtures of them). PR boxes could then be viewed as the unit of non-local correlation, in analogy with the bit, qubit and ebit, which are the units of classical and quantum information theoretic resources.

\paragraph*{Interior points.}
We have only considered conversions between extremal probability
distributions.  It would be interesting to consider the interior
points of the polytope, which comprise quantum correlations. In particular we would like to find out if distillation of such mixed correlations is possible, i.e., if given a number of copies of a mixed box we can by local operations obtain some number of extremal boxes. Note that Cirel'son's bound \cite{Cirelson} shows that the quantum correlations $\mathcal{Q}$, are a proper subset of the set of all non-signalling correlations $\mathcal{P}$. Thus it is impossible to distill correlations in $\mathcal{Q}$ to extremal correlations. But apart from this, we do not know of any constraint on possible distillation of non-local correlations. 

Finally, one could consider distillation in a new context, where we allow some communication between the parties but account for it at the end of the protocol (as noted above, an analogous approach was considered in Ref.~\cite{ieee2} in the context of classical distillation of shared randomness). Alternatively, following Ref.~\cite{collinspopescu}, one could introduce a new element, that of secrecy. Suppose that inputs and outputs are considered to be secret, and that Alice and Bob have a supply of noisy (that is non-extremal) boxes. Can Alice and Bob distill a supply of extremal boxes, whose inputs and outputs are also secret, via public communication? 

As we outlined in the introduction, non-local extremal correlations can be a very powerful resource for communication complexity problems. This will also be the case for correlations that can be distilled to these with no or little communication. On the other hand, Cirel'son's bound and results in communication complexity \cite{cleve} put limits on the power of quantum mechanics as a resource in distributed tasks. A better understanding of the possible interconversions between non-local correlations might bring an information theoretic explanation of these limitations.

\acknowledgments
We would like to thank Wim van Dam, Andreas Winter and Harry Buhrman for useful discussions. We acknowledge financial support from the Communaut\'{e} Fran\c{c}aise de Belgique under grant ARC 00/05-251, from the IUAP programme of the Belgian government under grant V-18, and from the EU through project RESQ (IST-2001-37559).


\end{document}